\documentclass{emulateapj}
\usepackage{apjfonts}

\newcommand{\zetavec}{\mbox{\boldmath $\zeta$}}

\lefthead{HAN} 
\righthead{WIDE-SEPARATION PLANETARY LENSING}

\begin{document}
\title{Characterization of Microlensing Planets with Moderately 
Wide Separations}

\author{Cheongho Han}
\affil{Department of Physics, 
Chungbuk National University, Cheongju 361-763, Korea;
cheongho@astroph.chungbuk.ac.kr}


\begin{abstract}
In future high-cadence microlensing surveys, planets can be 
detected through a new channel of an independent event produced 
by the planet itself.  The two populations of planets to be 
detected through this channel are wide-separation planets and 
free-floating planets. Although they appear as similar short 
time-scale events, the two populations of planets are widely 
different in nature  and thus distinguishing them is important.
In this paper, we investigate the lensing properties of events 
produced by planets with moderately wide separations from host 
stars.  We find that the lensing behavior of these events is 
well described by the Chang-Refsdal lensing and the shear caused 
by the primary not only produces a caustic but also makes the 
magnification contour elongated along the primary-planet axis.  
The elongated magnification contour implies that the light 
curves of these planetary events are generally asymmetric and 
thus the asymmetry can be used to distinguish the events from 
those produced by free-floating planets.  The asymmetry can be 
noticed from the overall shape of the light curve and thus can 
hardly be missed unlike the very short-duration central perturbation 
caused by the caustic.  In addition, the asymmetry occurs 
regardless of the event magnification and thus the bound nature 
of the planet can be identified for majority of these events.  
The close approximation of the lensing light curve to that of 
the Chang-Refsdal lensing implies that the analysis of the light 
curve yields only the information about the projected separation 
between the host star and the planet.
\end{abstract}

\keywords{gravitational lensing -- planetary systems}


\section{Introduction}

The microlensing signal of a planet is a short-term perturbation to 
the smooth standard single-lens light curve of the primary-induced 
lensing event occurring on a background source star.  For the 
detections of short-duration microlensing signals of planets, 
current lensing searches are being conducted in combination of 
survey and the follow-up observations,  where the survey observations 
aim to maximize the number of detections of lensing events by 
monitoring a large area of sky and follow-up observations are 
focused on intensive monitoring of the events detected by the 
survey observations.  Under this strategy, events can be followed-up 
while they are undergoing lensing magnification and thus only planets 
within a certain range, so called the lensing zone \citep{gould92}, 
can be detected.

If the monitoring frequency of survey observations is sufficiently 
high, planets can be detected through a channel of an independent 
event produced by the planet itself.  Currently, the MOA collaboration 
\citep{bond02} is devoting a portion of observation time for 
high-frequency survey by using its telescope with a $2.18\ {\rm deg}^2$ 
field of view.  It is also planned to upgrade the telescope for an 
even wider field of view of $4.0\ {\rm deg}^2$ (F.\ Abe 2009, private 
communication ).  In 2009, the OGLE collaboration \citep{udalski03} 
plans to increase the field of view of the camera from $0.4\ {\rm deg}^2$ 
to $1.4\ {\rm deg}^2$ for higher monitoring frequency (A.\ Udalski 
2009, private communication).  KMTNet (Korea Microlensing Telescope 
Network) is a planned experiment to achieve continuous resolution 
of planetary lensing signals of down to sub-Earth-mass planets by 
using a network of three 1.6 m telescopes to be located in Chile, 
South Africa, and Australia (B.\ Park 2009, private communication).  
Together with a $4.0\ {\rm deg}^2$ field of view of each telescope, 
the experiment will monitor the same field in every 10 minutes.  
{\it MPF} (Microlensing Planet Finder) is a space-based experiment 
proposed to NASA to achieve a similar cadence to KMTNet 
\citep{bennett09}.  With the enhanced monitoring frequency of these 
experiments, the efficiency of planet detection will greatly improve.

The two populations of planets to be additionally detected from 
high-frequency surveys are wide-separation planets and free-floating 
planets \citep{han07}.  Although appearing as similar short time-scale 
events, the two populations of planets are widely different in nature.  
Therefore, it is important to distinguish the two populations.  In 
this paper, we investigate the characteristics of events produced by 
bound planets with wide separations and examine the feasibility of 
distinguishing them from those produced by free-floating planets.

The paper is organized as follows.  In \S\ 2, we briefly describe 
the lensing properties of planetary events.  In \S\ 3, we investigate 
the lensing properties of events produced by planets with moderately 
wide separations and examine these properties whether they can be used 
for distinguishing the planets from free-floating planets.  We also 
investigate what kind of information about wide-separation planets
can be obtained from the analysis of light curves.  In \S\ 4, we 
compare the proposed method with other methods.  We summarize and 
conclude in \S\ 5.

\begin{figure*}[th]
\epsscale{0.9}
\plotone{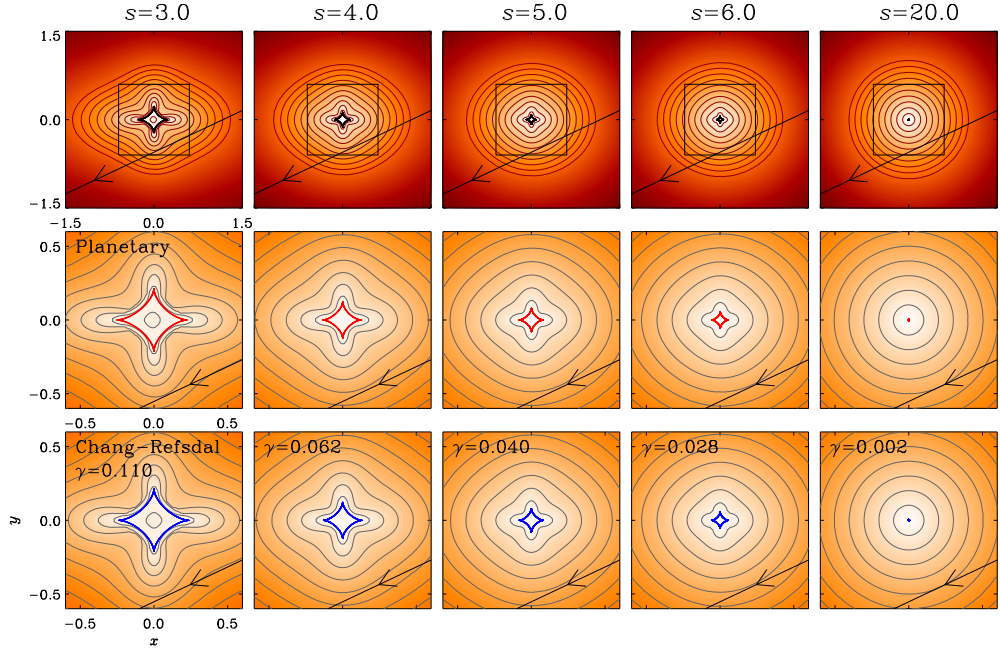}
\caption{\label{fig:one}
Maps of magnification pattern of wide-separation planets with 
various separations from the host star.  The separation $s$ is 
normalized by the Einstein radius corresponding to the mass of 
the primary, $\theta_{\rm E}$.  The planets have a common mass 
ratio of $q=6 \times 10^{-3}$ and the source star has a radius 
of $0.003 \theta_{\rm E}$ or equivalently $0.04\theta_{\rm E,p}$, 
where $\theta_{\rm E,p}$ is the Einstein radius corresponding to 
the planet mass.  For each map, the temperature scale represents 
the magnification where brighter tone implies higher magnification.  
Contours are drawn at the levels of the single-lens magnifications 
corresponding to the normalized lens-source separations of $u=0.1$, 
0.2, 0.3, $\cdot\cdot \cdot$, 1.0, respectively.  The upper panels 
show the regions around the planetary caustic  with $|\xi_{p}|\leq 
1.5$ and $|\eta_{\rm p}| \leq 1.5$, where $(\xi_{\rm p},\eta_{\rm p})$ 
are the coordinates centered at the center of the planetary caustic 
and lengths are normalized by $\theta_{\rm E,p}$.  The panels in 
the middle row show the blowups of the box region in the corresponding 
upper panels.  The panels in the bottom row show the maps constructed 
based on Chang-Refsdal lensing.
}\end{figure*}

\section{Planetary Lensing}

The planetary lensing behavior is described by the formalism of 
binary lensing with a small mass-ratio companion.  For binary 
lensing, the equation of lens mapping from the lens plane to the 
source plane is expressed as \citep{mao91}
\begin{equation}
\zeta=z-{1\over \bar{z}}-\left({q\over 1-q} \right){1\over \bar{z}-\bar{z}_{\rm p}},
\label{eq1}
\end{equation}
where $\zeta=\xi+i\eta$ and $z=x+iy$ represent the complex 
notations of the source and the image positions, respectively, 
the primary is at the center, $z_{\rm p}$ represents the position
of the planet, $\bar{z}$ denotes the complex conjugate of $z$, 
and $q$ represents the planet/primary mass ratio.  Here all 
lengths are normalized by the Einstein radius corresponding 
to the total mass of the lens system, $\theta_{\rm E}$.  One 
important characteristic of binary lensing is the formation of 
caustics, which represent the set of source positions on which 
the lensing magnification of a point source becomes infinite.  
As a result, the magnification pattern in the region around the 
caustic deviates from the smoothly varying pattern in the region 
away from the caustic and thus events resulting from source 
trajectories passing close to the caustic produce noticeable 
perturbations in lensing light curves.  The set of caustics 
forms closed curves, each of which is composed of concave 
curves that meet at cusps.  For planetary lensing, there exist 
two sets of caustics.  One tiny ``central caustic'' is located 
close to the primary lens.  The other ``planetary caustic'' with 
a relatively large size is located away from the primary at the 
position of 
\begin{equation}
\zetavec_{\rm c}={\bf s}(1-s^{-2}),
\label{eq2}
\end{equation}
where {\bf s} is the normalized separation vector to the position 
of the planet from the position of the primary.  The size of the 
planetary caustic as measured by the separation between the two 
cusps on the star-planet axis is \citep{han06a}
\begin{equation}
\Delta\xi_{\rm c}\simeq {4q^{1/2}\over s(s^2-1)^{1/2}}.
\label{eq3}
\end{equation}
For a wide-separation planet, the size of the caustic decreases 
as $\Delta\xi_c\propto s^{-2}$.

If a lensing event is produced by the approach of a source 
trajectory close only to the magnification region around the 
planetary caustic induced by a wide-separation planet, the 
resulting light curve, to the first order of approximation, 
appears as a single-lens light curve produced by the planet 
itself.  For a planet with a moderately wide separation, however, 
the magnification pattern exhibits deviations from single lensing 
due to the lensing influence of the primary.  For planets 
with separations in this regime, the magnification pattern is 
approximated by the Chang-Refsdal lensing.  The Chang-Refsdal 
lensing represents single lensing superposed on a uniform 
background shear \citep{chang79,chang84,mao92,an06}.  The lens-mapping 
equation of the Chang-Refsdal lensing is represented by
\begin{equation}
\zeta=z-{1\over \bar{z}}+\gamma\bar{z}, 
\label{eq4}
\end{equation}
where $\gamma$ is the shear.  For the case of the planetary 
lensing with a wide-separation planet, the shear exerted by 
the primary lens on the magnification region of the planet is
\begin{equation}
\gamma\sim {m_\star/m_{\rm p}\over \hat{s}^2}=
{1\over (1+q)s^2}\sim {1\over s^2},
\label{eq5}
\end{equation}
where 
$m_\star$ and $m_{\rm p}$ are the masses of the primary star and the 
planet, respectively, and $\hat{s}=[(1+q)/q]^{1/2}s\sim q^{-1/2}s$ 
is the planetary separation normalized by the Einstein radius 
corresponding to the mass of the planet, $\theta_{\rm E,p}$. The 
shear induces a single caustic that forms around the lens.  The 
caustic has the shape of a hypocycloid with four cusps.  The 
size of the caustic as measured by the separation between the 
two confronting cusps is 
\begin{equation}
\Delta\xi_{{\rm c},{\rm C-R}}\sim 4\gamma\sim 4s^{-2},
\label{eq6}
\end{equation}
where the caustic size is normalized by $\theta_{\rm E,p}$.
The shear decreases rapidly with the increase of the planetary 
separation.  As a result, for an event produced by a very 
wide-separation planet, it becomes difficult to distinguish 
the resulting light curve from that of an event produced by a 
free-floating planet.

\begin{figure*}[th]
\epsscale{0.9}
\plotone{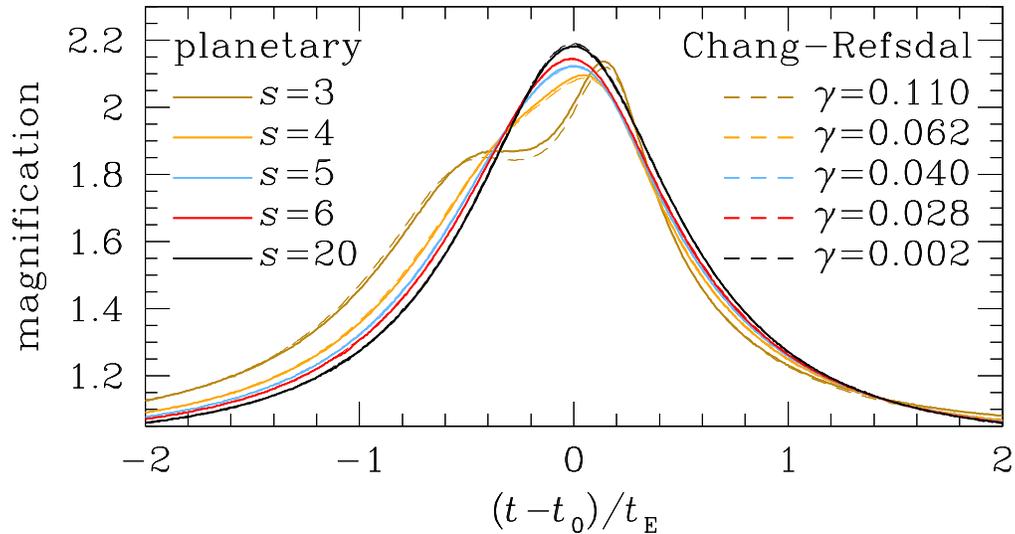}
\caption{\label{fig:two}
Light curve of events produced by wide-separation planets with 
various separations from the host star.  The source trajectory 
responsible for the individual light curves are marked in the 
corresponding panels in Fig.~\ref{fig:one}.  For each color, the 
solid curves is based on exact planetary lensing while the dashed 
curve is based on Chang-Refsdal lensing.
Note that the time scale $t_{\rm E}$ corresponds to the planetary mass.
}\end{figure*}

\section{Lensing by Wide-Separation Planets}

To investigate the lensing properties of events produced by 
wide-separation planets, we produce maps of magnification pattern
in the region around the planets.  For events produced by small-mass 
planets, the size of the source star is not negligible compared to 
the Einstein radius of the planet and thus finite-source effect 
is important.  We  take the effect into consideration by using the 
ray-shooting method.  In this method, a large number of light rays 
are uniformly shot from the observer plane through the lens plane 
and then collected in the source plane.  Then, the lensing magnification 
considering the finite-source effect is computed by comparing the 
number density of rays within the source radius with the density 
on the lens plane.  In addition, this method is easily applicable 
to any form of lensing as long as the mapping equation is known, 
and thus allows to study both planetary and Chang-Refsdal lensing 
with minor modification of the computation code.

Figure~\ref{fig:one} shows the constructed maps of magnification-pattern 
for planets with various separations from the host star.  For each map, 
the temperature scale represents the magnification where brighter tone 
implies higher magnification.  Contours are drawn at the levels of the 
single-lens magnifications corresponding to the normalized lens-source 
separations of $u=0.1$, 0.2, 0.3, $\cdot\cdot\cdot$, 1.0, respectively, 
i.e., $A(u)=(u^2+2) /u(u^2+4)^{1/2}$.  The panels in the upper row show 
the regions around the planetary caustic with $|\xi_{\rm p}|\leq 1.5$ 
and $|\eta_{\rm p}|\leq 1.5$, where $(\xi_{\rm p},\eta_{\rm p})$ are 
the coordinates centered at the center of the planetary caustic, 
$\xi_{\rm p}$ is aligned with the star-planet axis, and lengths are 
normalized by $\theta_{\rm E,p}$.  The planets have a common mass 
ratio of $q=6\times 10^{-3}$.  We assume a ratio of the source radius 
$\theta_\star$ to the Einstein radius of the primary lens of $\theta_\star
/\theta_{\rm E}=0.003$, which is a typical value for Galactic bulge events 
occurring on a bright main-sequence star.  When normalized by the Einstein 
radius of the planet, this corresponds to $\theta_\star/\theta_{\rm E,p}
\sim 0.04$.  The maps in the second row show the blowups of the box regions  
in the corresponding upper panels.  The panels in the lower row show the 
maps constructed by using the mapping equation of the Chang-Refsdal lensing.

From the investigation of the maps, we find the following properties
of the magnification pattern.
\begin{enumerate}
\item
First, the caustic occupies a small portion of the area enclosed 
by the Einstein ring of the planet.
\item
Second, the Chang-Refsdal lensing is a good approximation for planets 
with moderately wide separations of $s\gtrsim 3$.
\item
Third, the shear caused by the primary star not only induces a 
caustic but also makes the magnification contour elongated 
along the star-planet axis.
\end{enumerate}

The properties of wide-separation planetary events have several 
important implications in characterizing planets to be detected 
through the channel of independent events in future lensing surveys.
First, the small size of the caustic implies that the bound nature 
of a planet can be identified from the caustic-induced perturbation 
\citep{han03} only for a small fraction of high-magnification events 
with source trajectories approaching close to the central caustic.  
If it is assumed that the perturbation region extends up to twice 
of the caustic size, the chance to be perturbed by the caustic is 
$\sim 2\times (\Delta\xi_{\rm c, C-R}/2)\sim 4s^{-2}$ among the 
events produced by the entrance of the source star within the 
Einstein ring of the planet.  This corresponds to $\sim 16\%$ 
of all events produced by a planet with $s=5$.

Second, the elongation of the magnification contour implies that 
the light curves of events produced by planets with moderately
wide separations are in general {\it asymmetric} \citep{night08}
and thus the 
asymmetry of the light curve can be used to identify the bound 
nature of the planet.  Figure~\ref{fig:two} shows several example 
light curves resulting from the source trajectories marked in the 
corresponding panels in Figure~\ref{fig:one}.  The perturbation 
induced by the caustic lasts only for a very short duration 
considering that the time scale of the planetary event itself is 
short and the perturbation takes up only a small fraction of the 
time scale.  This implies that the perturbation can often be missed 
even in high-frequency surveys.  On the other hand, the asymmetry 
of the light curve can be noticed from the overall shape of the light 
curve and thus can hardly be missed.  In addition, the asymmetry 
occurs regardless of the event magnification and thus bound planets 
can be identified for majority of events.  Although depending on the 
assumed photometric precision and the type of events, we find that 
the asymmetry can be noticed for planets with separations up to 
$s\sim 10$.  Considering that the physical Einstein radius of a 
typical Galactic lensing event is $r_{\rm E}\sim 2$ AU, this 
corresponds to a physical separation of 20 AU, which encompasses 
all planets except Neptune in our Solar System.

Third, the fact that the lensing behavior of a wide-separation 
planet is well described by the Chang-Refsdal lensing implies 
that the analysis of the light curve yields only the information 
about the projected separation $s$ and the constraint on the mass 
ratio $q$ is poor. This is because the Chang-Refsdal lensing is 
described by the single parameter $\gamma$ and it is related to 
only the projected separation by $\gamma\sim s^{-2}$.

\section{Discussion}

Besides the methods of using the central perturbation and the 
asymmetry of the light curve, wide-separation planets can also 
be distinguished from free-floating planets by using other 
methods.  One method is using an additional bump in the lensing 
light curve produced by the primary star.  This method is 
applicable to events where the source trajectory passes the 
magnification region of the planet also approaches the 
magnification zone of the primary \citep{distefano99}. The 
disadvantage of this method is that only a fraction of events 
produce bumps and the fraction decreases with the increase of 
the star-planet separation.

Another photometric method is analyzing the baseline flux of the 
event.  If non-zero baseline flux is measured and the source of 
the baseline flux is identified as the lens itself, the planet 
can be identified as a bound planet. The source of the baseline 
flux can be identified from high-resolution images obtained by 
using instruments such as the {\it Hubble Space Telescope}, the 
Very Large Telescope (VLT), or the Keck Telescopes.  However, 
the applicability of this method is limited by the telescope time 
for high-resolution follow-up observations.

Finally, the two populations of planets can also be distinguished 
by conducting astrometric follow-up observations.  Due to the much 
longer range of astrometric lensing effect than photometric effect 
combined with the much larger mass of the primary than the planet, 
the primary star will affect the centroid motion of the source star 
induced by the planet \citep{han06b}. This method requires astrometric 
observations by using high-precision interferometers such as those 
to be mounted on space-based platforms, e.g., the {\it Space 
Interferometry Mission}, or on very large ground-based telescopes, 
e.g., VLT or Keck. Therefore, the application of this method is 
also limited by the telescope time for follow-up observations.

\section{Conclusion}

We investigated the lensing properties of events produced by 
the approach of the source trajectory only to the magnification 
region of a planet with a moderately wide separation from the 
host star.  We found that the lensing behavior of these events
is well described by the Chang-Refsdal lensing and the shear 
caused by the primary not only produces a caustic but also makes 
the magnification contour elongated along the primary-planet 
axis.  The elongated shear implies that the light curves of 
these planetary events are generally asymmetric and thus the 
asymmetry can be used to distinguish the events from those 
produced by free-floating planets.  We, therefore, propose to 
model the light curves of short time-scale events with either 
planetary or Chang-Refsdal lensing even though the light curves 
seemingly appear to be single-lensing ones.  The asymmetry can be 
noticed from the overall shape of the light curve and thus can 
hardly be missed unlike the very short-duration central perturbation 
caused by the caustic.  In addition, the asymmetry occurs regardless 
of the event magnification and thus the bound nature of the planet 
can be identified for majority of events.  The close approximation 
of the lensing light curve to that of the Chang-Refsdal lensing 
implies that the analysis of the light curve yields only the 
information about the projected separation between the host star 
and the planet.

\acknowledgments 
This work is supported by Creative Research Initiative program
(2009-008561) of Korea Science and Engineering Foundation.

\end{document}